# LUMINESCENT AND ABSORPTIVE METAL-COATED EMULSIONS FOR MICRO-VELOCIMETRY

O. Mesdjian[1,2,3], Y. Chen[1,2,3], J. Fattaccioli[1,2,3]*

[1] Ecole Normale Supérieure - PSL Research University, Département de Chimie, 24 rue Lhomond, F-75005 Paris, France

[2] Sorbonne Universités, UPMC Univ. Paris 06, PASTEUR, F-75005, Paris, France.

[3] CNRS, UMR 8640 PASTEUR, F-75005, Paris, France

*Corresponding author e-mail: jacques.fattaccioli@ens.fr

## ABSTRACT

Fluorescent latex beads have been widely used as tracers in microfluidics over for the last decades. They have the advantages to be density matched with water and to be easily localizable using fluorescence microscopy. We have recently synthesized silver-coated oil droplets that are both luminescent and absorptive. They have a mean diameter of 6 µm and a their density has been matched to the density of water by adjusting the thickness of the metallic layer. In this work we used these particles as tracers to measure the velocity profile of a water flow in a PDMS microchannel with a rectangular cross-section, which allowed us to confirm the predictions of the Stokes equation with results comparable to those of common submicronic polystyrene particles.

# INTRODUCTION

Measuring the local instantaneous velocity of a liquid at micron scale has been an important challenge these last decades [1,2] to understand e.g. the mixing properties of microfluidic devices [3] or the no-slipping condition commonly assumed in hydrodynamics [4]. With the development of lab-on-a-chip applications, micro-velocimetry has been used to characterize e.g. electrokinetic [5], *in vivo* [6], or biomedical science [7] related flows. For a pressure driven flow in a microchannel, the mean flux can be determined by weighing the liquid going out from a device during a certain amount of time but this technique does not allow measuring the spatial and temporal variations of the velocity inside the channel. To address this issue, micro-particle image velocimetry (µPIV) or micro-particle tracking velocimetry (µPTV) have been developed [8] to measure the spatiotemporal features of complex fluid flows. They both rely on the observation of moving particles and calculation of the relative displacement of the particles during a finite amount of time. Commonly used tracers [9] in micro-velocimetry experiments are fluorescent polystyrene beads with diameters ranging from 0.2 µm to several microns [8]. Using fluorescence microscopy, we can increase the contrast between the particles and the surrounding fluid, making the tracking fast and easy.

We recently developed [10] a fabrication protocol of metal-coated particles with a liquid core that are both luminescent and light absorptive, making them interesting candidates for multimodal observation in micro-velocimetry. The synthesis of this type of particles is relatively easy by using soybean oil-in-water emulsion droplets, i.e. a metastable liquid-liquid suspension. The oil droplets are first coated with a polydopamine shell [11] synthesized *in-situ* and then a thin silver layer by electroless plating. Both diameter of the droplets and the thickness of the silver shell can be adjusted to obtain particles with a density close to water, as the weights of oil and silver compensate.

In this work, we use both brightfield and epifluorescence microscopy to measure the velocity profile of a pressure-driven flow of an aqueous solution seeded with metal-coated droplets in a PDMS microchannel of rectangular cross section. Using these particles as µPTV tracers allows us to confirm the predictions of the Stokes equation.

MATERIALS AND METHODS

**Materials**: The Poloxamer 188 block-polymeric surfactant ($HO(C_2H_4O)_{79}$-$(C_3H_6O)_{28}$-$(C_2H_4O)_{79}H$) was kindly provided by Croda France SAS. The sodium alginate was purchased from Sigma-Aldrich. Ultrapure water (Millipore, 18.2 $M\Omega.cm^{-1}$) was used for all experiments. All the chemicals used for the metallized emulsion preparation were purchased from Sigma-Aldrich. Fluorescent polystyrene beads (DragonGreen, diameter of 0.51 µm) were purchased from Polyscience.

**Emulsion Fabrication:** We first disperse by manually stirring 15 g of soybean oil in an aqueous phase containing 2.5 g of a surfactant (Poloxamer F-68, initial proportion of 30 %w/w) and 2.5 g of a thickening agent (sodium alginate, initial proportion of 4 %w/w). This crude, polydisperse emulsion is further sheared and rendered quasi-monodisperse in a Couette cell apparatus under a controlled shear rate (5000 $s^{-1}$), following the method developed by Mason et al. [12]. Before decantation, the emulsion is diluted in order to have a proportion of 1% w/w of Poloxamer F-68 and 5% w/w of oil. After one night of decantation, the oil phase is get, diluted with a solution of Poloxamer F-68 with an initial proportion of 1% w/w. After several decantation steps to remove very small droplets, the emulsion (final proportion of 50% w/w of oil) is stored at 12°C in a Peltier-cooled cabinet.

**Fabrication of the metallodielectric emulsions**. The protocol is described on **Figure 1A** and is inspired from Nocera et al. [10]. The cream of the soybean oil emulsion is rinsed twice with a Tris buffer (pH = 8.5, 20 mM) supplemented with a surfactant (Tween 20, initial proportion of 0.2% w/w in the Tris buffer). To do this rinse, we mix 70 µL of the emulsion and 200 µL of the Tris-Tween solution, centrifuge during 30 s at 4000 rpm and remove the aqueous phase (200 µL) with a micro pipet. For the polydopamine coating, we disperse the 70 µL of the rinsed emulsion in 100 µL of a solution of dopamine (initial concentration of 5 mg.mL$^{-1}$ in the Tris-Tween solution) and we add 85 µL of the Tris-Tween solution. Then, we add 5 µL of potassium permanganate ($KMnO_4$, initial concentration of 40 mg.mL$^{-1}$ in water) in the solution to enable the transformation of dopamine in polydopamine at the surface of the droplets. The solution is stirred with a rotor (60 rpm) during two hours at room temperature in obscurity. To remove the unreacted molecules the sample is rinsed twice, as described before, using 180 µL of the Tris-Tween solution. Then we coat the droplets with a silver layer using an electroless plating process. To do this, we first disperse the droplets in a solution of silver nitrate ($AgNO_3$, initial concentration of 18 mg.mL$^{-1}$ in the Tris–Tween solution) and stir the obtained solution on a rotor (60 rpm) during 90 minutes at room temperature in the obscurity. In 50 µL of the obtained dispersion, we add 150 µL of a solution of ascorbic acid. The acid ascorbic solution has an initial concentration of 15.5 mg.mL$^{-1}$ in the Tris-Tween solution supplemented with a surfactant (polyvinylpyrrolidone, initial proportion of 0.02% w/w in the Tris-Tween solution).

**Microfluidic device fabrication:** The devices are made in PDMS (polydimethylsiloxane), using the standard soft lithography techniques [13]. In brief, we fabricate SU-8 (SU-8 3050, Microchem) masters on a silicon wafer, then proceed to PDMS (RTV 615, 1:10 ratio for the reticulating agent, RTV 615, Momentive Performance Materials) molding and thermal curing at 80 °C during two hours. We treat the PDMS surfaces and the glass coverslip (VWR, 50×24 mm) that closes the channel with air plasma (300 mTorr, 40 s) before sealing both parts of the chip together. The microfluidic system is then physically connected with small tubing to two small glass beakers used to impose a hydrostatic pressure difference to the flow. The channel dimensions are: width ($w$) = 200 µm, height ($h$) = 65-70 µm and length ($L$) = 2 cm.

**Microscopy.** Brightfield and fluorescent images particles were acquired on a Zeiss Axio Observer Z1 microscope (Oberkochen, Germany) connected to a Flash 2.8 sCMOS camera (Hamamatsu Photonics, Japan). Epi-illumination was done with a HXP 120L metal-halide lamp and a GFP filter set (Excitation wavelength : 475 nm, Emission wavelength : 530 nm). All images were taken with a 100x oil immersion objective (NA: 1.25, DOF ≤ 1 µm). Exposure times were respectively set to 50 ms and 80 ms for in brightfield and Epifluorescence condition.

**Numerical simulations and Image analysis.** All computations have been performed with Mathworks Matlab software.

## RESULTS AND DISCUSSION

**Metal-coated luminescent and absorptive oil droplets.** The droplets of an oil-in-water emulsion are coated first with polydopamine and then with silver (**Figure 1A**) following a fabrication protocol we developed recently [10]. The naked and monodisperse soybean oil droplets are first dispersed in an oxidative and alkaline aqueous solution of dopamine. The dopamine self-polymerizes at the surface of the droplets following a reaction scheme that involves the formation of an intermediary indole form [11]. After the addition of silver nitrate, silver ions are adsorbed on the polydopamine layer and can be used as a seeding layer for a further metallization. The addition of ascorbic acid, a reducing agent, and PVP, a stabilizing polymer, leads to silver-coated emulsion droplets. As compared to Nocera et al. [10], we adjusted in this work the silver nitrate concentration to obtain metallized particles with a density close to the density of the aqueous phase, as indicated by the fact that particles were neither moving upwards nor downwards for several hours in the aqueous solution they were suspended into.

Silver-coated droplets displacements are recorded using brightfield (**Figure 1B** on the left) and epifluorescence (**Figure 1B** on the right) microscopy. On brightfield images, the particles appear black on a white background due to the presence of the absorptive metal-layer, whereas on fluorescence images, they appear white on a black background as we have shown before that silver-coated emulsion droplets have luminescent properties. In both cases, they are easily localizable by microscopy.

The silver-coated droplet population has a diameter, measured by microscopy, equal to 6.5 ± 1.5 μm (see the distribution of diameter in **Figure 1C**). After fabrication, the final suspension has a volume concentration of 0.3 %, a viscosity similar to water and is used for micro-velocimetry experiments without any additional rinsing step.

**Experimental setup.** The experimental setup is illustrated in **Figure 2**. For the velocimetry experiments, we fabricated a PDMS-on-glass microfluidic device using classical soft lithography techniques (see the **Materials and Methods** section and [13]). The tubing and the channel are filled with the solution containing the particles. The flow rate is controlled by imposing a defined hydrostatic pressure difference between the inlet and the outlet of the channel, the hydrodynamic resistance of the tubing being negligible compared to the one of the channel. The pressure drop $\Delta P$ between the two reservoirs is set by a difference in height $\Delta h$ controlled by a precision translation stage. (0.02 mm accuracy). A typical pressure range is $\Delta P = \rho g \Delta h = 1 - 10 \, Pa$ (where $\rho$ is the density of water and $g$ the gravity) obtained with $\Delta h$ = 0.1-1 mm and corresponding to a maximal velocity of the order of 100 μm.s$^{-1}$. The evaporation of the liquid in the reservoirs is negligible. The decrease of the flow rate due to liquid level in the beakers changing during the experiment is negligible. The chip is put on a brightfield and epifluorescence inverted microscope and image series are recorded with an sCMOS camera.

**Flow in a rectangular microchannel.** For the general case of a channel with a rectangular cross-section, the axial velocity in the (*y*,*z*) plane of **Figure 2** is known analytically in terms of a Fourier series [14]:

$$V(y,z) = \frac{\Delta P}{2\mu L}\left\{\left[\left(\frac{h}{2}\right)^2 - z^2\right] - \sum_{n=0}^{\infty} a_n \cos\left(\frac{(2n+1)\pi}{2}\cdot\frac{z}{h/2}\right)\cosh\left(\frac{(2n+1)\pi}{2}\cdot\frac{y}{h/2}\right)\right\} \quad (1)$$

where $\Delta P$ is the pressure drop between the two ends of the channel and $\mu$ is the dynamic viscosity of the suspension (taken equal to water as the suspension is diluted). In no-slip boundary conditions, the coefficients $a_n$ writes as:

$$a_n = \frac{(-1)^n h^2}{\left[\frac{(2n+1)\pi}{2}\right]^3 \cosh\left(\frac{(2n+1)\pi}{2}\cdot\frac{w}{h}\right)} \quad (2)$$

To simplify the analysis and write the velocity $V$ as a function of $z$ alone, we assume that the microchannel height $h$ is negligible compared to its width $w$, so that the problem is the same as the unidirectional pressure-driven flow in a Hele-Shaw chamber. In these conditions, the axial velocity in the channel $V$ is governed by the Stokes equation that writes as:

$$\frac{\partial P}{\partial x} = \mu \frac{\partial^2 V}{\partial z^2} \quad (3)$$

The integration of equation (3), with a no-slip condition at the walls ($V = 0$ at $z = 0$ and $z = h$), gives a parabolic relation for $V(z)$ which writes as:

$$V(z) = \frac{\Delta P}{2\mu L} z(h-z) \quad (4)$$

From the equation ( 4 ) we deduce the maximum axial velocity $V_{max}$ of the particules in the microchannel :

$$V_{max} = \frac{h^2}{8\mu L}\Delta P = \frac{3}{2whR}\Delta P \qquad (5)$$

Where

$$R = \frac{12\mu L}{wh^3} \qquad (6)$$

corresponds to the hydrodynamic resistance of the channel. With the dimensions of our channel, we find $R = 0.63$ mbar.min.µL$^{-1}$.

The **Figure 3A** shows the 3D axial velocity profile calculated from (1) and for the specific case corresponding to our channel dimensions, for a pressure drop $\Delta P = 3$ Pa. From this graph, we plot the axial velocity profile in the center plane y = 0 (**Figure 3B**, solid line), that is compared to the parabolic velocity profile of **Equation** ( 4 ) (dotted line). The difference of maximal velocity $V_{max}$ is about 2%, which justifies a parabolic profile approximation.

**Velocity profile measurement.** The dimensions of the observation window of the microscope ($f_x \times f_y = 70 \times 50$ µm$^2$) are set by the magnification of the objective and the size of the CMOS sensor of the camera. As shown in **Figure 4A**, we chose to put the observation window in the center of the channel as the axial velocity of the particles does not vary so much in this area. **Figure 3C** shows indeed that in the central part of the profile that corresponds to the width of the window $f_y$, the variation of the velocity is about 2% from its maximal value. We can thus consider that all the particles within the observation window have the same axial velocity.

According to the law of refraction [15], the z position of the focal plane in the channel filled with water is proportional to the height of the objective, directly read on the microscope, and writes as:

$$z_{FP} = z_{Obj} \cdot \frac{n_{water}}{n_{oil}} \tag{7}$$

where $z_{FP}$ is the vertical position of the focal plane, $z_{Obj}$ the vertical position of the objective, $n_{water}$ and $n_{oil}$ the respective index of water and oil ($n_{oil} = 1.5$).

The objective is moved over height intervals $\Delta z_{Obj} = 5$ µm (corresponding to $\Delta z_{FP} = 4.4$ µm for the focal plane) from the bottom of the channel to the top to acquire the full variation of the axial velocity profile. We localize the bottom of the channel using adsorbed impurities on the glass coverslip. At each step, an image sequence long enough to observe the displacement of several particles in the observation field is recorded with the camera. For each time-lapse recording, the instantaneous velocity $V(z)$ of the particles is then determined by measuring the axial displacement $\Delta x$ of each particle within a short timeframe $\Delta t$ set by the camera.

The **Figure 4** shows the two pictures chosen in brightfield (**B**) and in fluorescence (**C**) to measure the displacement $\Delta x$ corresponding to a timeframe $\Delta t$. For each height $z$, we measure the velocity of N = 4 particles from which we calculate the mean and the standard deviation of $V(z)$.

**Velocity profiles acquired with polystyrene beads and metal-coated droplet.** We first validate our experimental method with fluorescent polystyrene beads (diameter : 0.5 µm, concentration : 0.1 % w/w) that are commonly used for micro-velocimetry. We also measured the velocity profiles of flowing metal-coated particles in brightfield and in fluorescence microscopy. The velocity profiles obtained with these two types of particles are shown in **Figure 5A, B** and **C** with $\Delta P = 3$ Pa and they are properly fitted by the parabolic function of the **Equation** ( 4 **)** with the pressure drop $\Delta P$ as a single adjustable parameter, $h$ being constant and equal to the height of the microchannel. Fitting parameters (see **Table 1)** are in agreement with the experimental pressure drop and the height of the microchannels, taking into account that the microchannel used for the measurement in epifluorescence condition had a smaller height than the ones used for PS beads and metal-coated droplets in brightfield condition ($h$ = 66 μm instead of $h$ = 68 μm as reported in **Table 1**).

**Measurement of the hydrodynamic resistance of the microchannels.** As expressed in **Equation (5)**, the maximum axial velocity $V_{max}$ depends linearly on the pressure drop $\Delta P$, the slope being inversely proportional to the hydrodynamic resistance $R$. We experimentally measured the relationship between the maximum velocity $V_{max}$ and the pressure drop $\Delta P$ by setting the focal plane at the height $z = h/2$ and measuring the maximum velocity for pressure drops ranging from 1 to 10 Pa. **Figure 5D,E and F** show the maximal velocity $V_{max}$ plotted as a function of $\Delta P$ for 0.5 μm polystyrene beads (**D**), silver-coated particles in brightfield (**E**) and in fluorescence (**F**) microscopy follow. As expected the relation is linear for the three experiments.

We derive the hydrodynamic resistance from the slope of the linear fits using the **Equation (6)** that gives the slope in function of the resistance $R$, the width of the channel $w$ and its height $h$. This resistance is compared to the expected resistance ($R$ = 0.63 mbar.min.μL$^{-1}$), using for $h$, $w$ and $L$ the dimensions of the channel measured by microscopy. The resistance values reported in **Table 2** are consistent considering the error bars. $R$ being proportional to $1/h^3$ (**Equation 6**), its value is very sensitive to variations in thickness of the channels, which explains why the hydrodynamic resistance found with metal-coated droplets in epifluorescence conditions is higher (0.66 mbar.min.μL$^{-1}$ compared to 0.60 mbar.min.μL$^{-1}$) than for PS bead and metal-coated droplets in brightfield condition.

# CONCLUSION

The silver-coated droplets we fabricated have the advantage to be both absorptive and fluorescent, which makes them observable using both fluorescence and brightfield microscopy. The mean diameter of these particles is 6.5±1.5 μm. By adjusting the width of the silver shell, we could obtain particles with an effective material density matching water density. Measuring the velocity of these particles at different heights within the microchannel, we could be able accurately determine the velocity profile of a pressure driven flow. As expected, we found a parabolic distribution of the liquid velocities and a linear relationship between the maximum velocity and the pressure drop, confirming the predictions of Stokes equation in our system. Silver-coated droplets can be thus used as micro-velocimetry tracers for water or aqueous solutions in a variety of microfluidic experiments.

## ACKNOWLEDGEMENTS

This work has received support of "Institut Pierre-Gilles de Gennes" (Laboratoire d'excellence : ANR-10-LABX-31, "Investissements d'avenir" : ANR-10-IDEX-0001-02 PSL and Equipement d'excellence : ANR-10-EQPX-34). We thank Remi Dreyfus (COMPASS Laboratory) for his careful reading of the manuscript.

**FIGURES**

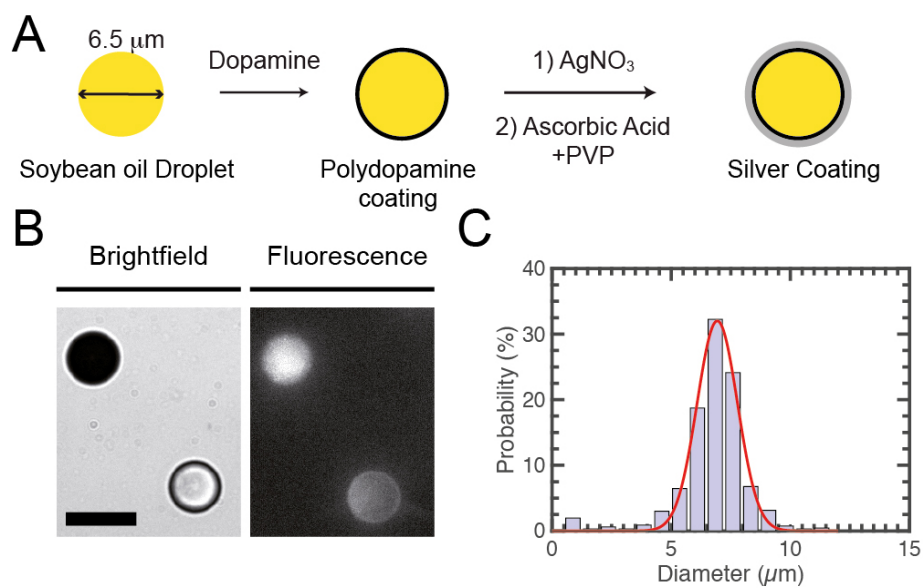

**Figure 1** : (A) Fabrication of metal-coated emulsion droplets. The naked soybean oil droplets are dispersed in an oxidative and alkaline aqueous solution of dopamine. Dopamine polymerizes at the surface of the droplets. After introducing silver nitrate, silver ions are adsorbed on the polydopamine layer. Addition of ascorbic acid, a reducing agent, and PVP, a stabilizing polymer, leads to silver-coated emulsion droplets. (B) Brightfield and fluorescence pictures of silver-coated droplets. (C) Size distribution histogram of the metal-coated emulsion. The diameter is equal to 6.5 ± 1.5 μm.

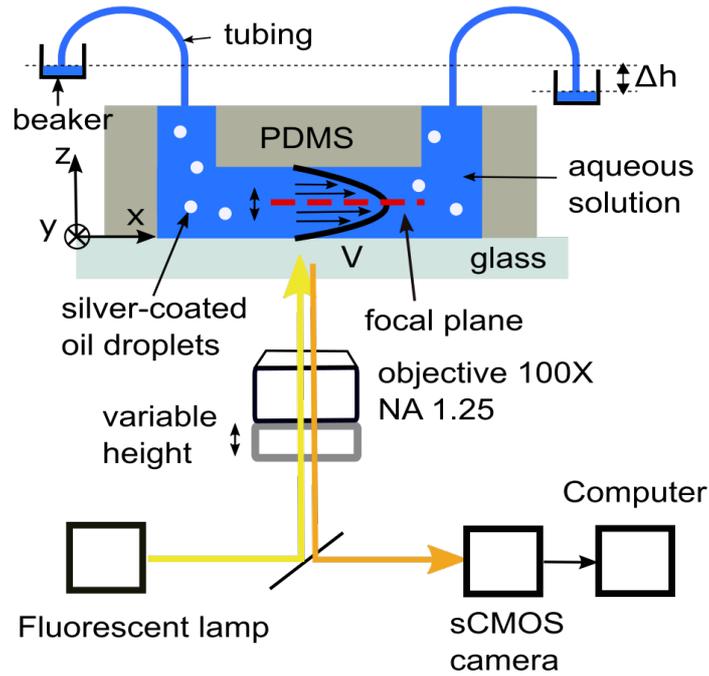

**Figure 2** : Schematic representation of the experimental setup. The microchannel (width $w$ = 200 μm, height $h$ = 65-70 μm and length $L$ = 2 cm ) containing the flowing particles is observed on a microscope in brightfield and epifluorescence conditions. The height difference $\Delta h$ between the two beakers gives rise to a hydrostatic pressure drop, $\Delta P = \rho g \Delta h$, between the two extremities of the channel. The pressure range $\Delta P$ varies between 1 to 10 Pa corresponding to a height range $\Delta h$ from 0.1 to 1 mm. The images of the particles in the channel are taken with an sCMOS camera and recorded on a computer.

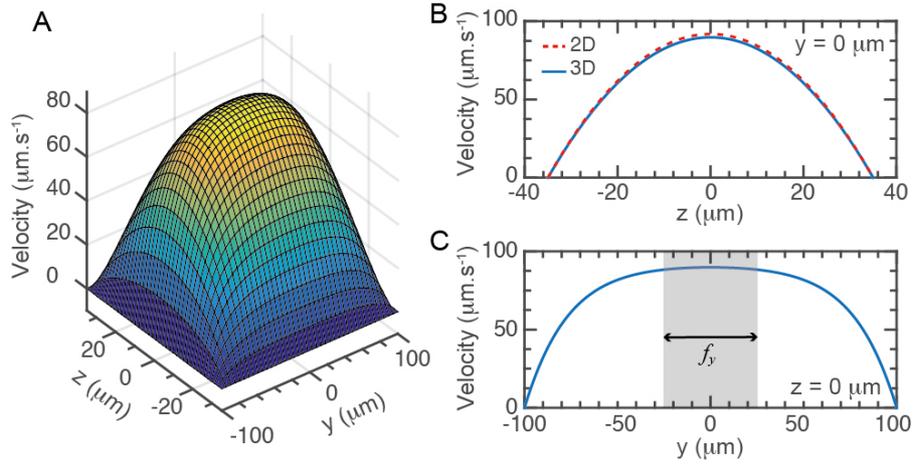

**Figure 3** : (A) Three-dimensional velocity profile in a microchannel ($n = 5$) with a rectangular cross-section and dimensions : width ($w$) = 200 µm, height ($h$) = 70 µm and length ($L$) = 2 cm. The pressure drop is 3 Pa. (B) The blue line represents the axial velocity distribution evaluated along the center-plane ($y = 0$). The red dotted line represents the parabolic profile chosen for the fits. (C) Axial velocity distribution evaluated along the center-plane ($z = 0$), when viewed from above. The central region of length $f_y$ = 50 µm corresponds to the width of the observation window along this direction. Over this length, the change of velocity is about 2% as compared to its maximal value.

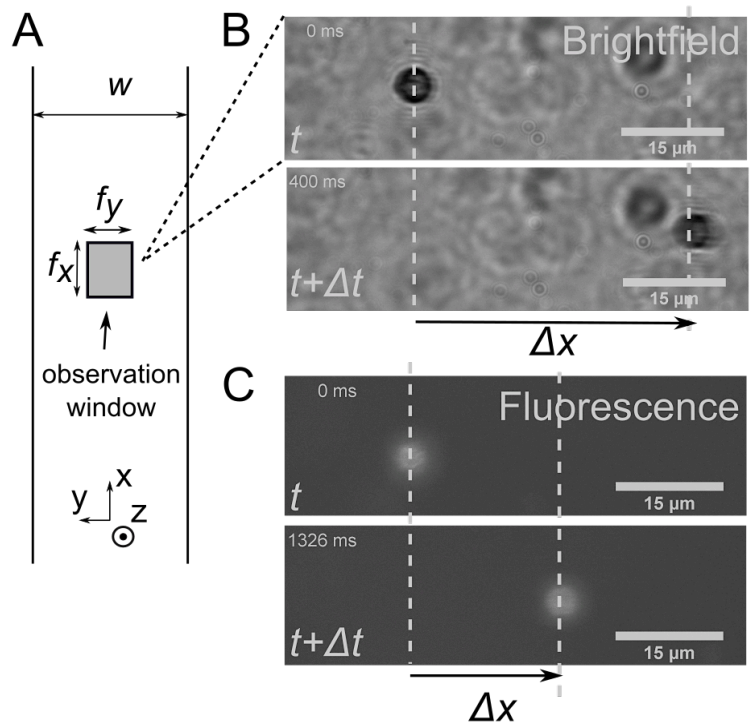

**Figure 4**: Schematic representation of the instantaneous axial velocity measurement method. (A) Schematic top-view of the channel and the positioning of the observation window (length $f_x$ = 70 μm and width $f_y$ = 50 μm) at the given height $z$ in the channel. The width of the channel is $w$ = 200 μm. (B,C) Two consecutive pictures of a metal-coated droplet in brightfield and in fluorescence microscopy. The displacement of the particle is denoted $\Delta x$ and the time between the two pictures is $\Delta t$.

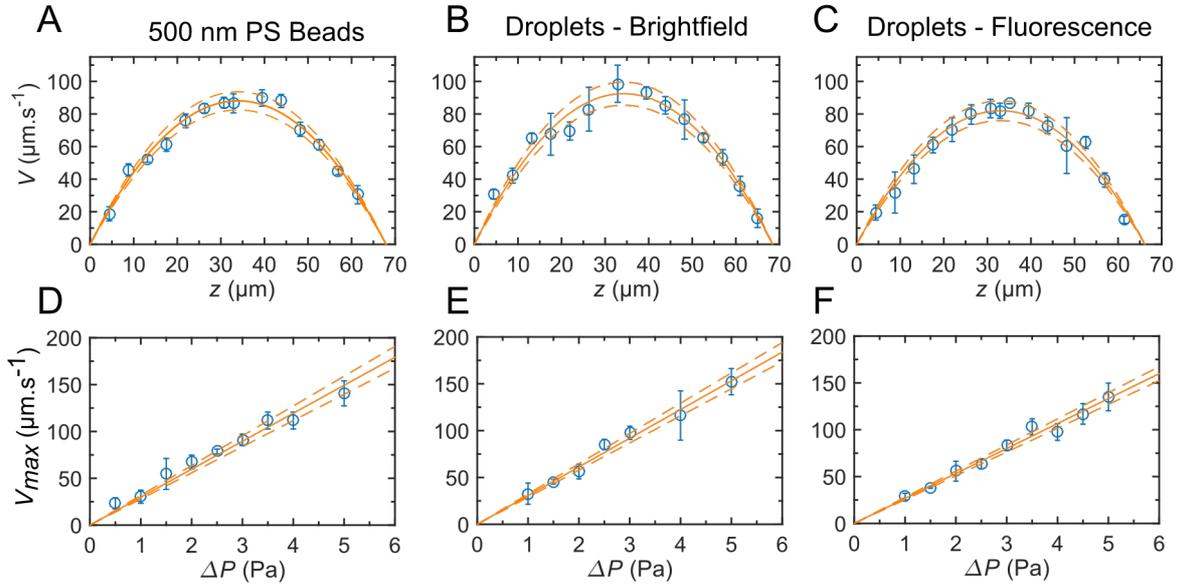

**Figure 5** : (A-C) Axial velocity *V* obtained by using 0.5 μm fluorescent polystyrene beads (A) and silver-coated droplets with bright field (B) and epifluorescence (C) measurement as a function of the height *z* in the microchannel for a pressure drop Δ*P* = 3 Pa. Data (blue open circles) are fitted by a parabolic profile with Δ*P* as the sole adjustable parameter. (D-F) Deduced maximal axial velocity $V_{max}$ from the velocity profile of A, B and C . Data (blue circles) follow a linear relationship. Dotted lines correspond to the axial velocity calculated for the maximal and minimal values extracted from the fit parameters. Experimental data were averaged on N=4 droplets and the error bar corresponds to the standard deviation of the instantaneous velocity values.

# TABLES

|  | ΔP (Pa) | Channel height $h$ (μm) |
|---|---|---|
| Experimental values | 2.9±0.1 | 68±2 |
| Polystyrene beads | 3.0±0.2 | 68±2 |
| Metal-coated droplets - Brightfield | 3.1±0.2 | 68±2 |
| Metal-coated droplets - Epifluorescence | 3.0±0.2 | 66±2 |

**Table 1:** Experimental and fitting parameters of the data and parabolic profiles shown in **Figure 5**. All the parameters have been determined using 0.5 μm fluorescent polystyrene beads and metal-coated droplets both in bright field and in fluorescence conditions. The pressure drop Δ$P$ and the height of the channel $h$ are extracted from the fits of the velocity profiles and are compared to the experimental pressure drop and the height of the channel measured by microscopy.

|  | $R$ (mbar.min.μL$^{-1}$) |
|---|---|
| Expected values from **Eq. (6)** | 0.63±0.03 |
| Polystyrene beads | 0.60±0.03 |
| Metal-coated droplets - Brightfield | 0.58±0.03 |
| Metal-coated droplets - Epifluorescence | 0.66±0.03 |

**Table 2:** Expected and calculated hydrodynamic resistances $R$ from the data shown on **Figure 5**. All the parameters have been determined for 0.5 μm fluorescent polystyrene beads and metal-coated droplets both in brightfield and in fluorescence conditions. The hydrodynamic resistance $R$ is derived from the slope of the fits of $V_{max}$ as a function of $\Delta P$ with the channel dimensions measured by microscopy.